\documentclass[10pt,twocolumn,english,superscriptaddress]{revtex4-1}
\usepackage[LGR,T1]{fontenc}
\usepackage[latin9]{inputenc}
\usepackage{geometry}
\usepackage{units}
\usepackage{amsmath}
\usepackage{graphicx}
\usepackage{color}
\makeatletter

\DeclareRobustCommand{\greektext}{%
  \fontencoding{LGR}\selectfont\def\encodingdefault{LGR}}
\DeclareRobustCommand{\textgreek}[1]{\leavevmode{\greektext #1}}
\DeclareFontEncoding{LGR}{}{}
\DeclareTextSymbol{\~}{LGR}{126}

\makeatother

\usepackage{babel}
\begin{document}

\title{Coherence time of over a second in a telecom-compatible quantum memory
storage material}

\author{Milo\v{s} Ran\v{c}i\'{c}}

\email{Corresponding author: (milos.rancic@anu.edu.au)}
\address{Centre for Quantum Computation and Communication Technology, Laser
Physics Centre, The Australian National University, Canberra, Australian
Capital Territory 0200, Australia. }

\author{Morgan P. Hedges}
\address{Physics, Princeton University, Princeton, New Jersey 08554, USA.}

\author{Rose L. Ahlefeldt}
\address{Centre for Quantum Computation and Communication Technology, Laser
Physics Centre, The Australian National University, Canberra, Australian
Capital Territory 0200, Australia. }

\author{Matthew J. Sellars}
\address{Centre for Quantum Computation and Communication Technology, Laser
Physics Centre, The Australian National University, Canberra, Australian
Capital Territory 0200, Australia. }

\maketitle

\textbf{
Quantum memories for light will be essential elements in future long-range quantum communication networks. These memories operate by reversibly mapping the quantum state of light onto the quantum transitions of a material system. For networks, the quantum coherence times of these transitions must be long compared to the network transmission times, approximately 100 ms for a global communication network. Due to a lack of a suitable storage material, a quantum memory that operates in the 1550 nm optical fiber communication band with a storage time greater than 1~us has not been demonstrated. Here we describe the spin dynamics of $\mathbf{^{167}Er^{3+}:Y_{2}SiO_{5}}$ in a high magnetic field and demonstrate that this material has the characteristics for a practical quantum memory in the 1550 nm communication band. We observe a hyperfine coherence time of 1.3 seconds. Further, we demonstrate efficient optical pumping of the entire ensemble into a single hyperfine state, the first such demonstration in a rare-earth system and a requirement for broadband spin-wave storage. With an absorption of 70 dB/cm at 1538 nm and \textgreek{L} transitions enabling spin-wave storage, this material is the first candidate identified for an efficient, broadband quantum memory at telecommunication wavelengths. 
}

Any future globally deployed quantum communication network \cite{Gisin2007,Kimble2008}
will require nodes connected by optical fiber. To minimize transmission
loss and maintain high data throughput, all elements of such a network,
particularly quantum repeater nodes, should transmit in one of the
low loss telecom bands for optical fiber at 1310 and 1550 nm. In its
simplest implementation, a quantum repeater relies on an efficient,
long-lived quantum memory \cite{Duan2001b}. 

Developing such a memory has proven very challenging. None of the proposed systems that operate directly in the low loss telecom band have the potential for long-term storage \cite{Riedinger2016, Saglamyurek2015}. For this reason, more complex ways of interfacing these candidate quantum memories with telecom are being investigated, including frequency conversion \cite{Radnaev2010, Dudin2010, Albrecht2014, Maring2014} or non-degenerate photon pairs \cite{Seri2017, Saglamyurek2011, Clausen2011, Bussieres2014, Zhang2016}. 

One of the most promising candidate memory systems is rare earth ions in solids. The potential for developing practical memories in these systems has been highlighted through a series of recent demonstrations
using non-Kramers ions (ions with an even number of electrons). Crystals
doped with either praseodymium or europium have demonstrated storage
on long lived spin states \cite{Ferguson2016,Jobez2015,Gundogan2015a},
multimode storage \cite{ Laplane2015,Ferguson2016}, large efficiencies \cite{Hedges2010,Sabooni2013}
and hyperfine coherence times of 6 hours using the Zero First Order
Zeeman (ZEFOZ) technique \cite{Zhong2015a}. These demonstrations were successful because the electronic angular momenta
of non-Kramers ions can be quenched by the crystal field for sites
with sufficiently low symmetry. Therefore, these ions can exhibit
the long ground-state hyperfine lifetimes necessary for efficient
optical spin pumping and long hyperfine coherence times. These properties
are key to the successful operation of efficient, long lived quantum
memories. Unfortunately for communication applications, none of the
non-Kramers ions have suitable optical transitions in any of the fiber
telecom bands. 

Compatibility with the telecom bands is offered by Kramers ions, with
an odd number of electrons. In particular, erbium has an optical transition
in the telecom band at 1538 nm. However, it is much more difficult
to make quantum memories with Kramers ions, and not a single Kramers
system has demonstrated an on-demand quantum memory. The root of the
difficulty is that, unlike for non-Kramers ions, the electronic magnetic
moment of Kramers ions cannot be quenched by a crystal field as they
possess a half-integer spin. For these ions there is a rapid electronic
spin relaxation which shortens the hyperfine state lifetimes. Baldit
et al. found the hyperfine state lifetimes to be limited to 100 ms
when demonstrating electromagnetic induced transparency in $^{167}\mbox{Er}^{3+}\mbox{:Y}_{2}\mbox{SiO}_{5}$
\cite{Baldit2010}. This is similar to the electron spin lifetime
\cite{Hastings-Simon2008} and only an order of magnitude longer than
the optical excited state lifetime, making efficient optical spin
pumping very difficult. While some other Kramers ions such as Nd offer
shorter optical lifetimes and so slightly improved spin pumping, the
electron-spin limit on the hyperfine lifetime is still very short
(100ms) preventing long term storage \cite{Usmani2010}. In comparison,
the hyperfine lifetimes for $\mbox{Pr}^{3+}\mbox{:Y}_{2}\mbox{SiO}_{5}$
and $\mbox{Eu}^{3+}\mbox{:Y}_{2}\mbox{SiO}_{5}$ are 5 minutes and
23 days respectively \cite{Holliday1993,Konz2003}.

Nevertheless, the advantage of direct telecom compatibility means
there has been considerable work towards developing quantum memories
using erbium. Instead of storing on spin states, this work has largely
focused on storing on the optical transitions. Atomic Frequency Comb
(AFC) delay lines have delayed quantum states of light in Er doped
glass fiber \cite{Saglamyurek2015}, however, the efficiency was limited
to \ensuremath{\sim} 1 \% and the storage time to 50 ns. Weak coherent
states have also been stored in crystals using a two-level Gradient
Echo Memory (GEM) technique with an efficiency of 0.25\%. The fidelity
of the recalled state in this demonstration was, however, well below
the non-classical limit \cite{Lauritzen2010}. For both these demonstrations
the inefficiency of the optical pumping of the ground state electron
spin levels was identified as limiting the memories' efficiency.
To avoid this limitation, new on-demand memory techniques not requiring
optical pumping to initialize the ensemble have been proposed \cite{Hetet2008,Dajczgewand2014}
and demonstrated \cite{Dajczgewand2014}. Although the latter technique
has shown significantly higher efficiencies, up to 40\% , the storage
times are still limited by the optical coherence time and quantum
storage has yet to be demonstrated. Currently, there is no proposed
strategy to achieve the fidelities and storage times required for
quantum repeater applications \cite{Razavi2009} without using long
lived hyperfine states.

Here, we investigate the hyperfine spin dynamics of $^{167}\mbox{Er}^{3+}\mbox{:Y}_{2}\mbox{SiO}_{5}$
in the presence of a large magnetic field, with the aim of increasing
both the hyperfine lifetime and coherence time to be comparable to
non-Kramers systems. As mentioned above, the hyperfine lifetime is
short because it is coupled to the electron spin, which itself flips
rapidly due to coupling to the lattice and to other electron spins.
In order to increase the hyperfine lifetime, therefore, it is necessary
to slow the electron spin flips. B\"{o}ttger et al. previously
demonstrated that  a  large magnetic field can
suppress electronic relaxation in $\mbox{Er}^{3+}\mbox{:Y}_{2}\mbox{SiO}_{5}$
\cite{Bottger2009}. The field decreases both the electronic spin-lattice
coupling and electronic cross-relaxation. The spin-lattice is turned
off because, with the electron spin splitting much larger than kT,
the energy density of available phonons is low (see Figure \ref{fig:T1 spin-spin}).
In other words, the electron spin is frozen into the lower spin state.
This also turns off the cross-relaxation, since there are no ions
in the upper spin state for the polarised ions to cross-relax with.
Although the spin is frozen, the lifetime of the excited electron
spin state is, in fact, shorter than at low field because of the large
density of empty phonon modes available for an excited spin to decay
into. This short lifetime has limited the usefulness of this technique
for previous memory demonstrations \cite{Lauritzen2010,Kurkin1980}.
However, these demonstrations worked with the $I=0$ Er isotopes,
and here we show that the situation is very different for $^{167}\mbox{Er}$
with $I=\nicefrac{7}{2}$. Once the electron spin is frozen in the
lower ground state, the hyperfine levels associated with this state
can have extremely long lifetimes and coherence times.

\begin{figure}[t]
\includegraphics{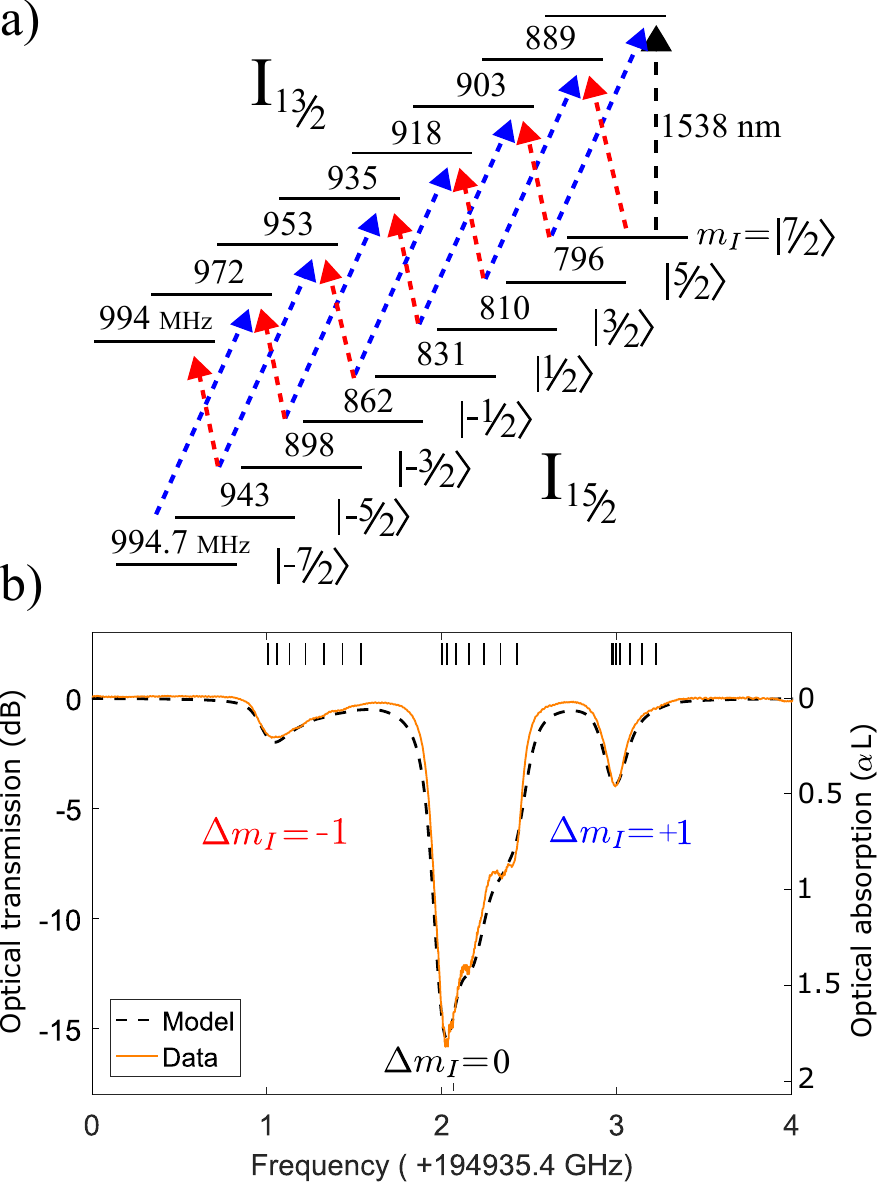}
\caption{\label{7 tesla absorbtion} 1538 nm optical transition of $^{167}\mbox{Er}^{3+}:\mbox{Y}_{2}\mbox{SiO}_{5}$  a) Energy level diagram for a field of 7~T.  The red-detuned $\Delta m_{I}=-1$ transitions are represented
by red arrows, and blue arrows for $\Delta m_{I}=+1$. The energy
spacing between the hyperfine states was determined by holeburning.
b) \textbf{\textit{Orange:}} Absorption spectrum at 1.4 K with a field
of 7 T along the $D_{1}$ axis.\textbf{\textit{ Black:}} model of
absorption based on holeburning measurements. \textbf{\textit{Vertical
Black Dashes:}} The centroids of the optical transitions used in the
black spectrum model.}
\end{figure}

The material used for this investigation was a 0.005\% doped $^{167}\mbox{Er}^{3+}\mbox{:Y}_{2}\mbox{SiO}_{5}$ crystal, enriched to 92\% isotopic purity. The Er ions substitute
for Y ions in two non-equivalent $C_{1}$ symmetry sites, and this
work used the site labelled `site 2' \cite{Guillot-Noel2006}. To most easily access the regime where the electron spin was frozen, the magnetic field was applied along a direction with a large ground state Zeeman splitting, the D$_1$ optical extinction axis (214 GHz/T).

The ground
state hyperfine structure was studied optically by exciting the $^{4}I_{15/2}\rightarrow{}^{4}I_{13/2}$
optical transition between the two spin-down projections of the lowest
crystal field levels, at a wavelength of 1538 nm (see supplementary materials for the energy level structure). In a field of 7
T, the Zeeman splitting of the electronic spin projections is more
than sufficient to resolve this transition. With a nuclear moment
of $I=\nicefrac{7}{2}$, $^{167}\mbox{Er}$ exhibits eight hyperfine
spin states $m_{I}=|\nicefrac{_{-}7}{2}\rangle,\cdot\cdot\cdot\,\,\cdot,|\nicefrac{_{+}7}{2}\rangle$
and the entire system studied comprises sixteen energy levels, as
shown Figure \ref{7 tesla absorbtion} a). The hyperfine energy spacings
were obtained by hole-burning measurements (see supplementary materials).

The absorption spectrum of the optical transition is shown in Figure
\ref{7 tesla absorbtion} b). Three bands are clearly resolved in
this spectrum, which we associate with the $\Delta m_{I}=-1,0$ and
1 transitions. The splitting between bands is 1 GHz, and they can
still be resolved at fields as low as 1 T. The absorption spectrum was fitted with a population model, shown in the figure, based on the measured energy level structure and including the contribution of $I=0$ impurity isotopes. The fitted lineshape was a Voigt profile with equal Gaussian and Lorentzian contributions and a linewidth of 150 MHz. The fit was used to determine the oscillator strengths  of each hyperfine transition.
Whilst the
strongest optical transitions have $\Delta m_{I}=0$, there is appreciable oscillator
strength in the fourteen $\Delta m_{I}=\pm1$ transitions. Relative
to the $\Delta m_{I}=0$ transitions, the oscillator strengths for
$\Delta m_{I}=-1\,(+1)$ start at approximately 25\% (31\%) for transitions
involving the $|\nicefrac{_{-}7}{2}\rangle$ states, and decrease
linearly to about 2.5\% (3.1\%) for the $|\nicefrac{_{+}7}{2}\rangle$
states. The decrease can be attributed to a difference in mixing between
hyperfine levels across the series. 

\begin{figure}[t]
\includegraphics{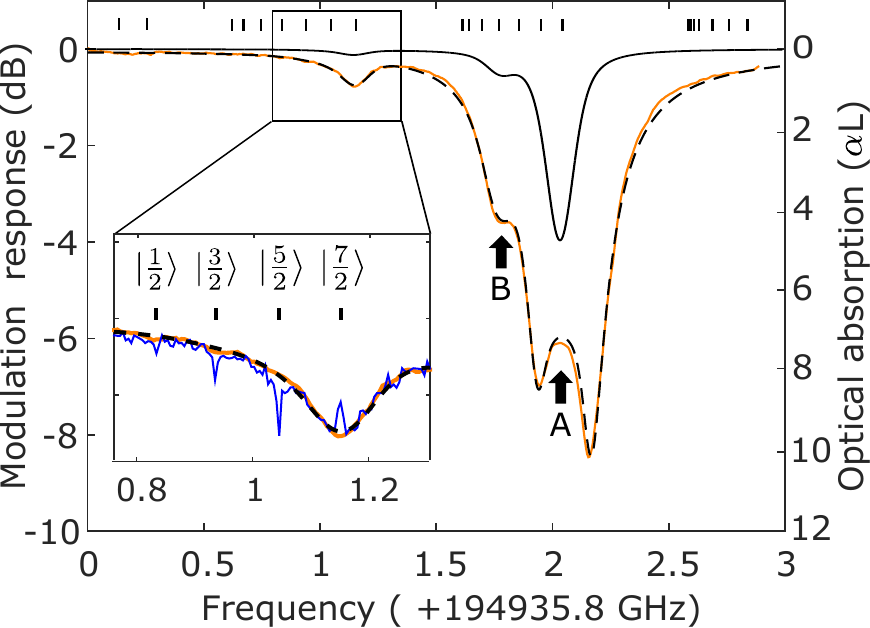}
\caption{\label{fig:Spin pumped}\textbf{\textit{ }}Absorption spectrum of
$^{167}\mbox{Er}^{3+}:\mbox{Y}_{2}\mbox{Si}\mbox{O}_{5}$ with 95\%
of the ensemble pumped into the $|\nicefrac{_{+}7}{2}\rangle$ hyperfine
spin state \textbf{\textit{(Orange}}, vertical axis on left), obtained by AM spectroscopy.
Pumping into the $m_{I}=|\nicefrac{_{+}7}{2}\rangle$ state was achieved
by sweeping the light over the set of $\Delta m_{I}=+1$ transitions,
corresponding to the 2.4 - 3.0 GHz frequency region.\textbf{\textit{
}}Arrow \textbf{A} indicates the center of the $|\nicefrac{_{+}7}{2}\rangle\rightarrow|\nicefrac{_{+}7}{2}\rangle$
absorption line. Arrow \textbf{B} indicates absorption of isotopic
impurities. \textbf{\textit{Black Solid Trace: }}Theoretical model
of optical absorption, corresponding to the vertical axis on the right.\textbf{\textit{
Black Dashed Trace:}} Theoretical model of the amplitude modulation
response. \textbf{\textit{Black Vertical Dashes:}} Centroids of inhomogeneously
broadened optical transitions, upon which the model is based, as well
as the last two $\Delta m_{I}=-2$ transitions on the far left. \textbf{\textit{Inset
(Orange):}} An expanded view of the $\Delta m_{I}=-1$ transitions.
\textbf{\textit{Inset (Dark Blue):}} Upon holeburning at the peak
absorption shown by arrow \textbf{A}, a hole and a series of anti-holes
are generated at the predicted energies. The labelled dashes indicate
the hyperfine ground state for each transition.}
\end{figure}

With the $\Delta m_{I}=-1$ and +1 bands clearly resolved, pumping
of the nuclear spin into a single state is possible through frequency
selection. By exciting only the seven $\Delta m_{I}=+1$ transitions
(the peak at 3 GHz in Figure \ref{7 tesla absorbtion} b), we were
able to pump $95\pm3\%$ of the $^{167}\mbox{Er}$ population into
the $|\nicefrac{_{+}7}{2}\rangle$ hyperfine ground state. This results
in a nearly eight-fold increase in the already high optical depth
of this transition. Such a high optical depth is difficult to measure
accurately with traditional absorption spectroscopy, and instead Amplitude
Modulation (AM) spectroscopy was used. This method is sensitive to
both the absorption and the phase shift of the transmitted light,
allowing a more accurate determination of the optical depth (for more
detail, refer to the Materials and Methods section). 

The resulting spectrum is shown in Figure \ref{fig:Spin pumped}.
Also shown in this figure is the model fitted to the spectrum to obtain
the optical depth, and the resulting absorption lineshape. This absorption
line has two components \textendash{} a shoulder at arrow B due to
the impurity isotopes, and a central peak at A due to ions in the
$m_{I}$= $|\nicefrac{_{+}7}{2}\rangle$ state. The AM spectrum for
this peak has a central dip due to the changing phase of the transmitted
light as the laser was swept through the highly absorbing peak. At
this frequency the absorption was determined to be $70\pm4$ dB/cm.

When hole-burning at the frequency of arrow A, the blue spectrum in
the inset of Figure \ref{fig:Spin pumped} is observed. The hole and
anti-holes demonstrate that spin-pumped population is shifted into
$|\nicefrac{_{+}5}{2}\rangle,|\nicefrac{_{+}3}{2}\rangle$ and $|\nicefrac{_{+}1}{2}\rangle$
hyperfine levels via the $\Delta m_{I}=-(1,2\,\&\,3)$ optical decay
paths.

The efficiency of the spin-pumping ($95\pm3\%$) was measured by the
fitting the model to only the $\Delta m_{I}=-1$ transitions, where
the decreasing trend in oscillator strengths and low absorption improve
the accuracy of the population estimate. The ability to polarize the
$^{167}\mbox{Er}$ ensemble into a single hyperfine state opens the
path towards several high-bandwidth memory techniques, which are discussed
later.

\begin{figure}[t]
\includegraphics{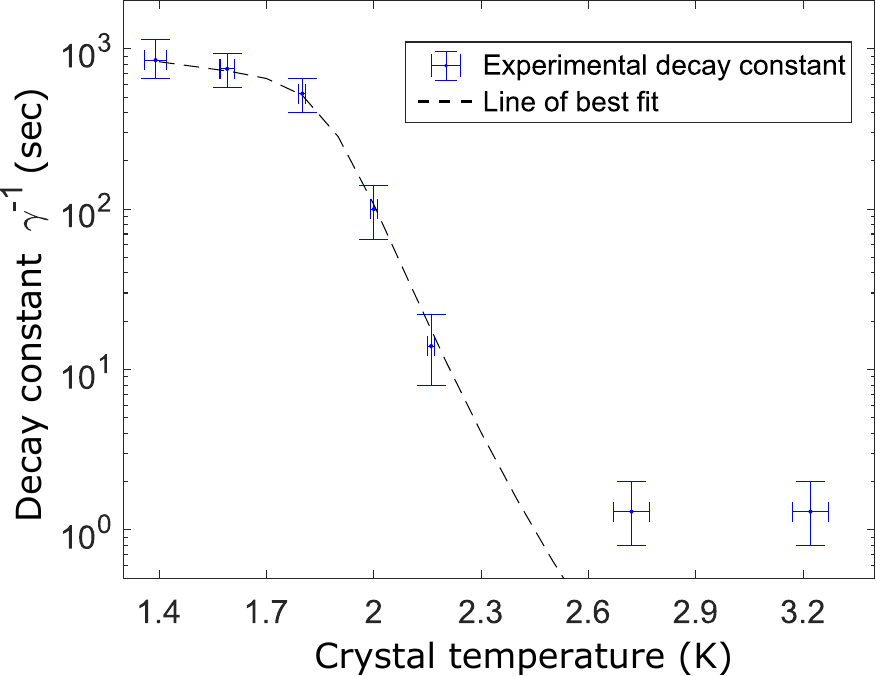}
\caption{\label{fig:Temp Dependance}\textbf{\textit{ }}The decay rate of $^{167}\mbox{Er}^{3+}$ nuclear spin polarization
as a function of temperature, in a field of 7 T. The value of $\gamma^{-1}$
is the hyperfine population decay rate $(T{}_{1})$. $x$-axis error bars show the maximum temperature variation during the measurement. 
$y$-axis error bars are given by the fit whose Root Mean Square Deviation (RMSD) is twice as large as the RMSD of the optimal fit to the spectrum.}
\end{figure}

Once the $^{167}$Er ensemble is pumped into the $|\nicefrac{_{+}7}{2}\rangle$
hyperfine state, the population slowly relaxes to thermal equilibrium
via spin-lattice coupling. Three main processes can contribute to
this relaxation: the one-phonon direct process, and the two-phonon
Raman and Orbach processes \cite{abragam1970}.

A general expression for the relaxation is:
\begin{alignat}{1}
\gamma(T) & =\gamma_{\,d}T+\gamma_{\,r}T^{9}+\gamma_{\,or}f^{3}\exp\left(-\frac{hf}{kT}\right)\label{eq:spinlattice}
\end{alignat}
where $f$ is the ground state electronic splitting.

To first order, these processes only cause decay through the $\Delta m_{I}=\pm1$
hyperfine transitions. This allowed us to model the population dynamics
as a series of coupled rate equations, with the same coupling rate
$\gamma$ between all adjacent hyperfine states (see supplementary
materials for more information). This population-dynamics model accurately
fit the recorded time series of spectra, hence supporting the assumptions
of the dynamics. 

The resulting time dependence of the spin-lattice relaxation rate
is shown in Figure \ref{fig:Temp Dependance}. The figure shows an
exponential increase in $T{}_{1}$ from 2.2 $\rightarrow$ 1.8 K.
This agrees with the exponential reduction in high energy phonons
associated with the Orbach process. Meanwhile, the plateau in transition
lifetimes below 1.8 K can be attributed to the direct phonon process. Fitting Equation \ref{eq:spinlattice}
to the low temperature data gave direct and Orbach coefficients of $\ensuremath{\gamma_{d}=9\times10^{-4}}~\mbox{s}\ensuremath{^{-1}}.\mbox{K}\ensuremath{^{-1}}$
and $\ensuremath{\gamma_{or}=8\times10^{-30}}~\mbox{s}\ensuremath{^{-1}}.\mbox{Hz}\ensuremath{^{-3}}$ respectively.
Above 2.6 K, the plateau in lifetimes indicates the presence of a
phonon bottleneck, limiting the rate of hyperfine relaxation \cite{abragam1970},
which was not included in the model. 

\begin{figure}
\includegraphics{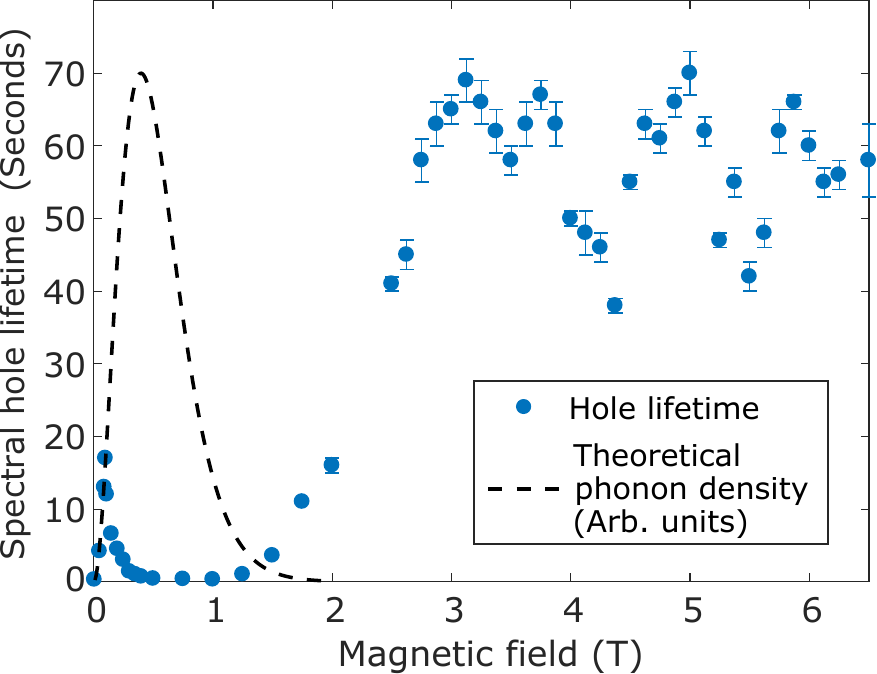}
\caption{\label{fig:T1 spin-spin}The lifetime of spectral holes burnt into the $\Delta m_{I}=+1$ absorption band of $^{167}\mbox{Er}^{3+}:\mbox{Y}_{2}\mbox{Si}\mbox{O}_{5}$ at 1.4 K, as a function of magnetic field along the $D_{1}$ axis. The theoretical phonon
density in black is a thermal Planck distribution. The curve assumes
the Zeeman splitting determined by Guillot-No\"el et al \cite{Guillot-Noel2006}.
$y$-axis error bars are the standard error in the fit to each point.}
\end{figure}

The other important decay mechanism in this system is hyperfine spin-spin
cross-relaxation. This process is masked in the previous measurement
because it does not redistribute the hyperfine populations, and therefore
does not change the population envelope. However, it will affect any
measurements operating on a sub-group of the entire ensemble, in particular
spectral holeburning, which is a
vital component of many quantum memory applications. Therefore, it is important
to characterize the cross-relaxation process.

We studied hyperfine cross-relaxation by measuring the lifetime of
a spectral hole burnt in the unpolarized spin ensemble as a function
of magnetic field. As shown in Figure \ref{fig:T1 spin-spin}, the
hole lifetime is short for fields below 3 T. In this regime, the electron
spin is not yet frozen and electron spin effects dominate the hole
lifetime. There is a small peak in the lifetime at 0.1 T, corresponding
to the field where the electronic Zeeman splitting is large enough
that electron spin cross-relaxation has been slowed, but where the
phonon energy density is still low. As the phonon density increases,
the lifetime drops, increasing only as the density drops back down
near zero at around 3 T. Above 3 T, hyperfine cross-relaxation dominates
the hole lifetime, leading to a maximum lifetime of about 70 s. The
hole lifetime does vary with field above 3 T, which is not expected
in a simple model of resonant hyperfine cross-relaxation. We attribute
the variation to the existence of fields where the detuning between
different Er hyperfine transitions can be bridged through hyperfine
cross-relaxation mediated by the high concentration of nuclear spin-$\nicefrac{1}{2}$ Y ions.

These measurements were performed in an unpolarized spin population. In a polarised system, faster cross
relaxation will occur for holes burnt into $\Delta m_{I}=\pm1$ states, but if population is transferred to states with $\Delta m_{I}=\pm2$ or higher the cross relaxation will be much slower than seen here.

We have shown that applying a large magnetic field freezes out the
electron spin dynamics. This results in a much quieter magnetic environment
in the crystal. Since magnetic field fluctuations typically dominate
the hyperfine coherence time, this should lead to a substantial increase
in the coherence time from its zero field value of 1 $\mu$s \cite{Baldit2010}. 

\begin{figure}[!t]
\includegraphics{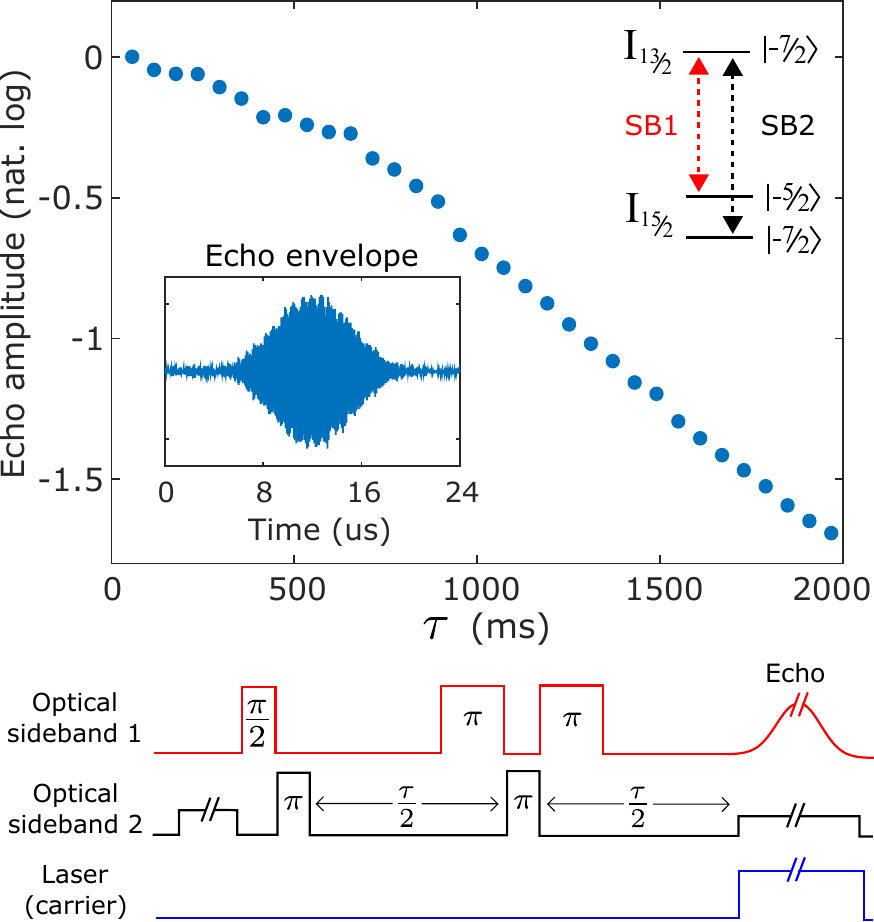}
\caption{\label{fig:Raman Echo} Raman echo measurement of $^{167}$Er coherence time. \textbf{\textit{Top: }} Normalized Raman echo
intensity as a function of total delay $\tau$, at 1.4 K and 7 T.\textbf{\textit{
Inset: }}The Raman echo time envelope with $\tau=60$ ms. \textbf{\textit{Bottom:}}
pulse sequence for Raman Echoes. The laser sits approximately 1 GHz
detuned from the lowest energy transitions, to prevent any absorption.
The two optical fields required to drive the ions were provided by
the sidebands of the same EOM used for AM spectroscopy, at 1155.3
MHz (SB1) and 2150 MHz (SB2). The carrier acted as a local oscillator
to optically detect the Raman echo. SB2 must be on for echoes to form,
as it transfers coherence to the optical transition. (Refer to Materials
and Methods for more detail). }
\end{figure}

We investigated the coherence time on the $|\nicefrac{_{-}7}{2}\rangle\leftrightarrow|\nicefrac{_{-}5}{2}\rangle$
ground state transition at 7 T using an all-optical Raman echo technique.
With this pulse sequence, coherence between hyperfine states is established
and manipulated via a common optical excited state, rather than by
direct magnetic resonance. The decay of the echo amplitude with time is shown in Figure
\ref{fig:Raman Echo}. Below 1 second, the decay is sub-exponential, and the $e^{-1}$ decay time is
$T_{2\,(echo)}=1300\pm10$ ms. 

With the electron spin frozen, this coherence time will be dominated
by the dynamics of the nuclear spins in the crystal, principally the
spin$-\nicefrac{1}{2}$ Y ions. These dynamics are substantially altered
from those in undoped $\mbox{Y}\ensuremath{_{2}}\mbox{SiO}\ensuremath{_{5}}$.
The large magnetic field of the Er ion creates a large frozen core
of Y spins whose frequencies are detuned from the bulk \cite{Bottger2006,Guillot-Noel2007}.
This means that the spin flips of those Y ions having the biggest
impact on the Er ion are slowed substantially, resulting in a much
longer coherence time than expected if all spins were flipping at
the bulk rate (3.6 ms, see supplementary materials for this calculation).
The spectral diffusion caused by the slow spin flips of the frozen
core also leads to the non-exponential shape to the echo observed
in Figure \ref{fig:Raman Echo}.

Here, we have shown that applying a large magnetic field of 7 T can
extend the hyperfine coherence time of $\mbox{Er}$ to 1.3 seconds.
In fact, Figure \ref{fig:T1 spin-spin} indicates that a similar result
could be achieved at fields as low as 3 T. 1.3 s
is the longest coherence time obtained in any Kramers system, and
is comparable to the longest coherence times obtained in non-Kramers
systems using the well-known ZEFOZ technique \cite{Zhong2015a}. From
a practical standpoint, the technique shown here has an advantage
over ZEFOZ, since it does not require precise alignment of the magnetic
field. This technique could be applied to achieve long coherence times in other Kramers ions considered for quantum memories such as Nd \cite{Usmani2010, wolfowicz2015, Clausen2011, Zhou2012}.

In the context of quantum repeater applications, the coherence time
seen here is already sufficient for a large scale network. Razavi
et al. show that a coherence time of 1 second is more
than sufficient for a 1000 km repeater network, even without error
correction \cite{Razavi2009}.

In addition to its long coherence time, $^{167}\mbox{Er}$:Y$_2$SiO$_5$ in the regime studied here has similar optical depth and optical pumping efficiency to  Pr:Y$_2$SiO$_5$, the non-Kramers material most widely used for quantum memory demonstrations. It also has 100 times larger hyperfine splittings than Pr, which means that it should have larger memory bandwidths and reduced noise from off-resonant excitation during the memory protocol. Collectively, the parameters presented here suggest $^{167}\mbox{Er}$ will rival, or exceed, the performance demonstrated by non-Kramers memories.

This system could also be considered for applications requiring long
term storage, i.e: a quantum `hard-drive'. For this application, it
should be possible to achieve a coherence time approaching the population
$T{}_{1}$ limit (10 minutes) at 1.6 K and 7 T using ZEFOZ. To achieve this longer coherence time it is necessary to turn off the hyperfine cross-relaxation. Since cross-relaxation only strongly couples states with $\Delta m_{I}=\pm1$, it can be eliminated by preparing storage qubits using non-adjacent hyperfine levels, for instance $|\nicefrac{_{-}7}{2}\rangle\leftrightarrow|\nicefrac{_{-}3}{2}\rangle$.

Finally, the ability to efficiently pump the $^{167}\mbox{Er}$ ensemble
into a single hyperfine state is crucial for high-bandwidth quantum
communication. It paves the way for broadband Raman memory techniques
which, until now, have been limited to atomic vapor systems \cite{Saunders2016}
and potentially nitrogen vacancies in diamond \cite{Poem2015}. Alternatively,
spin-pumping allows for potentially GHz-bandwidth spin-wave storage
using techniques already demonstrated in rare earth systems, such
as GEM \cite{Hedges2010} and drastically improves the efficiency
of AFC protocols \cite{DeRiedmatten2008}.

\subsection*{Materials and Methods:}

\textbf{Experimental Setup}\\
A 0.005\% doped $^{167}\mbox{Er}^{3+}\mbox{:Y}_{2}\mbox{SiO}_{5}$
$3\times4\times5$ mm ($D_{1}$,$D_{2}$,$b$) crystal provided by
Scientific Materials Corp (Bozeman, Montana) was maintained at 1.4
K in an Oxford helium bath cryostat with a 15 T super-conducting magnet (see Figure 1 of the supplementary materials).
Both the magnetic field and light propagating direction were parallel
with the $D_{1}$ optical extinction axis. As there was only one cryostat
window, the crystal was placed against a mirror, onto which the beam
was focused with a 70 $\mu$m waist. The reflection down the $D_{1}$
axis gave a total absorption length of 6mm. The optical transitions
were excited with a Thorlabs TLK-1550R laser, stabilized to a home-made
fiber reference cavity with a 1 second linewidth less than 100 kHz.
Intensity and frequency modulation of the light was achieved with
a 10 GHz JDSU AM-EOM and detected with a Lab Buddy 10 GHz detector.
\\
\\
\textbf{AM Spectroscopy}\\
This method was based on detecting the optical beat between the sidebands
and carrier of an EOM as a function of the modulation frequency. The
RF modulation sweep was generated by a spectrum analyzer, which was
also used for detection . This formed a closed modulation loop, as
shown in Figure 1 of the supplementary materials. The optical sidebands were weak
($\sim1\%)$ compared to the carrier power and the laser was kept
far detuned ($0.5-1$ GHz) from the absorption. 

This technique is phase sensitive, allowing for accurate measurement
of large optical depths, and as the laser remains locked to the reference
cavity spectra can be recorded in conjunction with precise optical
pulses and sweeps. 
\\
\\
\textbf{Energy Structure and Spin Pumping}\\
The hyperfine structure of the $I_{15/2}$ and $I_{13/2}$ states
was determined at 7 T and 1.4 K through a series of holeburning measurements.
A 0 to 4 GHz tunable RF source was used to generated a series of 20
spectral holes across 3 sets of measurements (see Supplementary Materials Section 3
for an example spectrum). Each spectrum was made by applying fixed frequency
RF for 100 ms. The absorption spectrum was recorded by a 0.01 to 2.9
GHz RF scan with the tracking generator and RF input on the spectrum
analyzer (50 ms scan, 1 MHz RBW). 
\\
\\
To spin pump the ensemble into either the $m_{I}=|\nicefrac{_{+}7}{2}\rangle$
or $|\nicefrac{_{-}7}{2}\rangle$ hyperfine ground state, a Voltage
Controlled Oscillator (VCO) was used to drive the EOM. A 3 second
saw-tooth scan (100 ms rep rate) was performed over the $\Delta m_{I}=-1$
or +1 absorption bands respectively, with $\sim$2 mW of optical power
in each sideband. The spin-pumped absorption spectrum was recorded
once again using AM spectroscopy. Repeating this with the laser (carrier)
frequency on either side of the absorption feature, it was possible
to stitch two scans together to span more than 2.9 GHz.
\\
\\
\textbf{Magnetic Field Dependence of Spectral Holes}\\
Spectral holes were burnt into the center of the $\Delta m_{I}=+1$
absorption band, as it afforded higher optical depth than the $\Delta m_{I}=-1$
band and deeper spectral holes than the $\Delta m_{I}=0$ band, which
had isotopic impurities and excessive optical depth. Between 0 and
1.5 T, spectral holes were generated using a tunable RF source. Above
1.5 T, 10 MHz wide spectral trenches were burnt instead, as there
was negligible spectral diffusion and long hole-burning lifetimes.
This was achieved by sweeping the spectrum analyzer tracking generator
output over 10 MHz for 10 seconds. The tracking generator output was
centered at approximately 1.5 GHz to keep the laser carrier and second
sideband off-resonant. The spectral features were measured using Phase
Modulation (PM) spectroscopy.
\\
\\
\textbf{Temperature Dependence of Spin Pumping}\\
In a 7 T field, spin pumping into the $m_{I}=|\nicefrac{_{+}7}{2}\rangle$
state was achieved using the same technique as described above. At
each temperature, 4 to 5 spectra were recorded at time intervals spaced
long enough to observe appreciable relaxation back to thermal equilibrium.
For example, at 1.6 K, spectra were recorded after 10 seconds and
(10, 20, 40, 80) minutes. See Section 5 of the Supplementary materials
for example spectra.
\\
\\
\textbf{Optical Raman Echoes}\\
Optical pulses ($\pi$ \& $\nicefrac{\pi}{2}$) were generated using
the EOM sidebands and two RF generators tuned to 1155.3 MHz and 2150
MHz. The pulse lengths were optimized using two pulse optical echoes
at either frequency, with 1.5 mW of optical power in each sideband.
For the $|I_{15/2},\nicefrac{_{-}5}{2}\rangle\leftrightarrow|I_{13/2},\nicefrac{_{-}7}{2}\rangle$
transition (1155.3 MHz), the optimal $\pi$-pulse length is $4\pm0.5\,\mu$s
and for the $|I_{15/2},\nicefrac{_{-}7}{2}\rangle\leftrightarrow|I_{13/2},\nicefrac{_{-}7}{2}\rangle$
a $\pi$-pulse was 1.5 $\mu s$ (2150 MHz). This pair of transitions
were chosen because they have the largest oscillator strengths, so
shorter pulses could be used. The hyperfine spin ensemble was prepared
by first pumping into the $m_{I}=|\nicefrac{_{-}7}{2}\rangle$ state
with a VCO for 10 seconds, as described above. This was followed by
applying a 100 $\mu$W pulse at 2150 MHz for 100 ms, to generate a
100 kHz wide absorption feature (anti-hole) in the $m_{I}=|\nicefrac{_{-}5}{2}\rangle$
subgroup. After the pulse sequence, the echo that reformed at the
wavelength of the 1150 MHz sideband was detected as a 1155.3 MHz optical
beat between the sideband and the carrier. This RF signal was then
mixed with a 1145.3 MHz RF source, and the 10 MHz beat signal was
recorded. Note: The detected echo amplitude in this technique depends
on the proportion of the initial ensemble transferred into the optical
$|-\nicefrac{7}{2}\rangle\leftrightarrow|-\nicefrac{7}{2}\rangle$
transition, which is determined by the probing power of sideband 2.

One advantage of this technique is that
the bandwidth of the $\sim1\,\mu s$ optical pulses used is much larger
than the inhomogeneous broadening of the hyperfine transition, and
so the width of the echo envelope gives a direct measure of this inhomogeneous
broadening. As shown in the inset to Figure 5, the echoes had a FWHM
of $7\pm1$ $\mu$s, corresponding to an inhomogeneous linewidth of
$130\pm20$ kHz.

\subsection*{Data availability}
The data that support the plots within this paper and other findings of this study are 
available from the corresponding author upon reasonable request.

\subsection*{Acknowledgments}
M. J. S would like to thank Charles Thiel for insightful discussions.
This work was supported by the Australian Research Council Centre
of Excellence for Quantum Computation and Communication Technology
(Grant No. CE110001027). M. J. S. was supported by an Australian Research
Council Future Fellowship (Grant No. FT110100919).

\subsection*{Author Contributions}
M. J. S and M. P. H. conceived the initial project. M. J. S, M. P.
H and M. R. designed the experimental setup. M. R. carried out the
experiment. M. R and R. L. A. analyzed the results. All authors contributed to writing the manuscript

\bibliographystyle{nature}

\end{document}